\documentclass[a4paper,11pt]{article}
\pdfoutput=1 % if your are submitting a pdflatex (i.e. if you have
\usepackage{jcappub} % for details on the use of the package, please
\usepackage{float}

\title{\boldmath Solving puzzles of GW150914  by primordial black holes}

\author[a,b,c,d,1]{S. Blinnikov,\note{Corresponding author.}}
\author[b,a]{A. Dolgov,}
\author[e,a]{N.K. Porayko} %\note{Also at Bonn MPI f. Radioastronomie.}}
\author[c,a,f]{and K.Postnov}

\affiliation[a]{ITEP,  Bol. Cheremushkinsaya ul., 25, 117218 Moscow, Russia}
\affiliation[b]{Novosibirsk State University,  Novosibirsk, 630090, Russia}
\affiliation[c]{Sternberg Astronomical Institute, Moscow M.V. Lomonosov State University,\\
 Universitetskij pr., 13,  Moscow 119234, Russia}
\affiliation[d]{Kavli IPMU (WPI), Tokyo University, Kashiwa, Chiba, Japan}
\affiliation[e]{Max-Planck-Institut f\"ur Radioastronomie, Bonn, Germany}
\affiliation[f]{Faculty of Physics, Moscow M. V. Lomonosov State University, 119991 Moscow, Russia}

\emailAdd{Sergei.Blinnikov@itep.ru}
\emailAdd{dolgov@fe.infn.it}
\emailAdd{nporayko@mpifr-bonn.mpg.de}
\emailAdd{kpostnov@gmail.com}

\abstract{The black hole binary properties inferred from the LIGO gravitational wave signal GW150914 posed
several serious problems. The high masses and low effective spin of black hole binary can be explained if they
are primordial (PBH) rather than the products of the stellar binary  evolution. Such PBH properties
are postulated ad hoc but not derived from fundamental theory. We show that the necessary features
of PBHs naturally follow from the slightly modified Affleck-Dine (AD) mechanism of baryogenesis.
The log-normal distribution of PBHs, predicted within the AD paradigm, is
adjusted to provide an abundant population of
low-spin stellar mass black holes. The same distribution gives a
sufficient number of quickly growing seeds of supermassive
black holes observed at high redshifts and may comprise an appreciable
fraction of Dark Matter which does not contradict any existing observational limits. Testable predictions of this
scenario are discussed.
}

\begin{document}
\maketitle
\flushbottom

\section{Introduction}
\label{sec:intro}

The recent discovery of gravitational waves (GWs) by LIGO~\cite{LIGO-PRL} not only opened a new window into high
redshift universe but simultaneously presented several problems.
The first observed signal GW150914 nicely fits
the hypothesis of the coalescence of two heavy black holes (BHs) which form a binary system
with masses $M_1= 36^{+5}_{-4} M_\odot$ and $M_2= 29^{+4}_{-4} M_\odot$. While the individual spins of coalescing BH
components are poorly constrained, the effective spin $\chi_{\rm eff}$ turns out to be close to zero, suggesting either low spins of the
components or their opposite direction, which is difficult to explain by the standard evolution from a massive binary system.
The parameter $\chi_{\rm eff}$ has been determined to be within $[0.19,0.17]$ from its effect on the inspiral rate
of the binary \cite{2016PhRvL.116x1102A}.
The final
BH mass is $M = 62^{+4}_{-4} M_\odot$ and its spin is substantial, $a = 0.65^{+0.05}_{-0.07}$, as expected from the
orbital angular momentum of the colliding BHs.

The first problem with GW150914 is a possible origin of so heavy BHs.
One of the proposed mechanisms involves
highly rotating massive (from $50 M_{\odot}$ up
to $100 M_{\odot}$) stars in close binary systems \cite{2016MNRAS.458.2634M, 2016A&A...588A..50M, 2016MNRAS.460.3545D}.
The rapid rotation of the stars, achieved due to strong tidal interaction,
provides effective mixing of nuclear burning products, thus avoiding significant post-main-sequence expansion and subsequent
mass exchange between the components.
To form so heavy BHs within this mechanism, the progenitors
should have low metal abundance to avoid heavy
mass loss during the evolution. Unfortunately, due to poor observational statistics, there is no strong observational confirmation
of the rates of binary systems, undergoing such chemically homogeneous evolution \cite{2016MNRAS.458.2634M}. Moreover, the LIGO
collaboration \cite{2016PhRvL.116x1102A} has shown that if both spins need to be aligned with orbital momentum, the individual spins with $90\%$ probability
have quite low values $a<0.3$. The second reliable LIGO detection, GW151226, turned out to be closer to the standard
binary BH system \cite{2016PhRvL.116x1103A}, which can be explained by classical binary system evolution.

The misaligned BH spins and low values of effective inspiral spin parameter $\chi_{\rm eff}$ detected in GW150914
may strongly constrain astrophysical formation from close binary systems \cite{2016MNRAS.462..844K}.
However, the dynamical formation of double massive BHs with misaligned spins
in dense stellar clusters is still possible \cite{2000ApJ...528L..17P, 2016ApJ...824L...8R}.

Mechanisms of BH formation from PopIII stars and subsequent formation of BH
binaries are analyzed in \cite{2016arXiv160404288D}.
The scenario proposed in \cite{2012ApJ...749...91F} is found to be the most appropriate.
However, the contribution of PopIII stars to the formation of BH binaries with masses
$\sim 30+30 M_\odot $
can be negligible, contradicting the population synthesis
results \cite{Kinugawa2014,Kinugawa2016}.

%Here we examine the origin of LIGO GW150914 event as primordial black holes in the framework of a specific
%model, a modified Affleck-Dine scenario, which was proposed as a viable
%mechanism of baryogenesis in the Universe.

A binary system of massive BHs like LIGO GW150914 could be formed from primordial black holes (PBH) \cite{2016PhRvL.116t1301B, 2016arXiv160308338S, 2016PASJ...68...66H}.
Here we propose a specific model of PBH formation which both naturally reproduce the extreme properties of
GW150914, the rate of binary BH merging events as inferred from the first LIGO science run 9-240
Gpc$^{-3}$~yr$^{-1}$ \cite{LIGO_doc2016},
%B. P. Abbott et al. (LIGO Scientific Collaboration and Virgo Collaboration), https://dcc.ligo.org/LIGO-P1600088/public/ main},
and provide seeds for early supermassive BH formation in galaxies.
The model is based on the modified Affleck-Dine~\cite{ad-bg} scenario for baryogenesis.
It was suggested  in 1994~\cite{ad-js}, discussed in more details in ref.~\cite{ad-mk-nk}, applied to an
explanation of the early supermassive BH (SMBH) observations at high redshifts~\cite{ad-sb}, and to a
prediction and study of the properties of possible antimatter stellar-like objects in the Galaxy~\cite{cb-ad,sb-ad-kp}.

\section{The model: Affleck-Dine scenario
\label{s-model}}

\begin{figure}[tbp]
\centering
% \begin{center}
\includegraphics[width=0.4\textwidth]{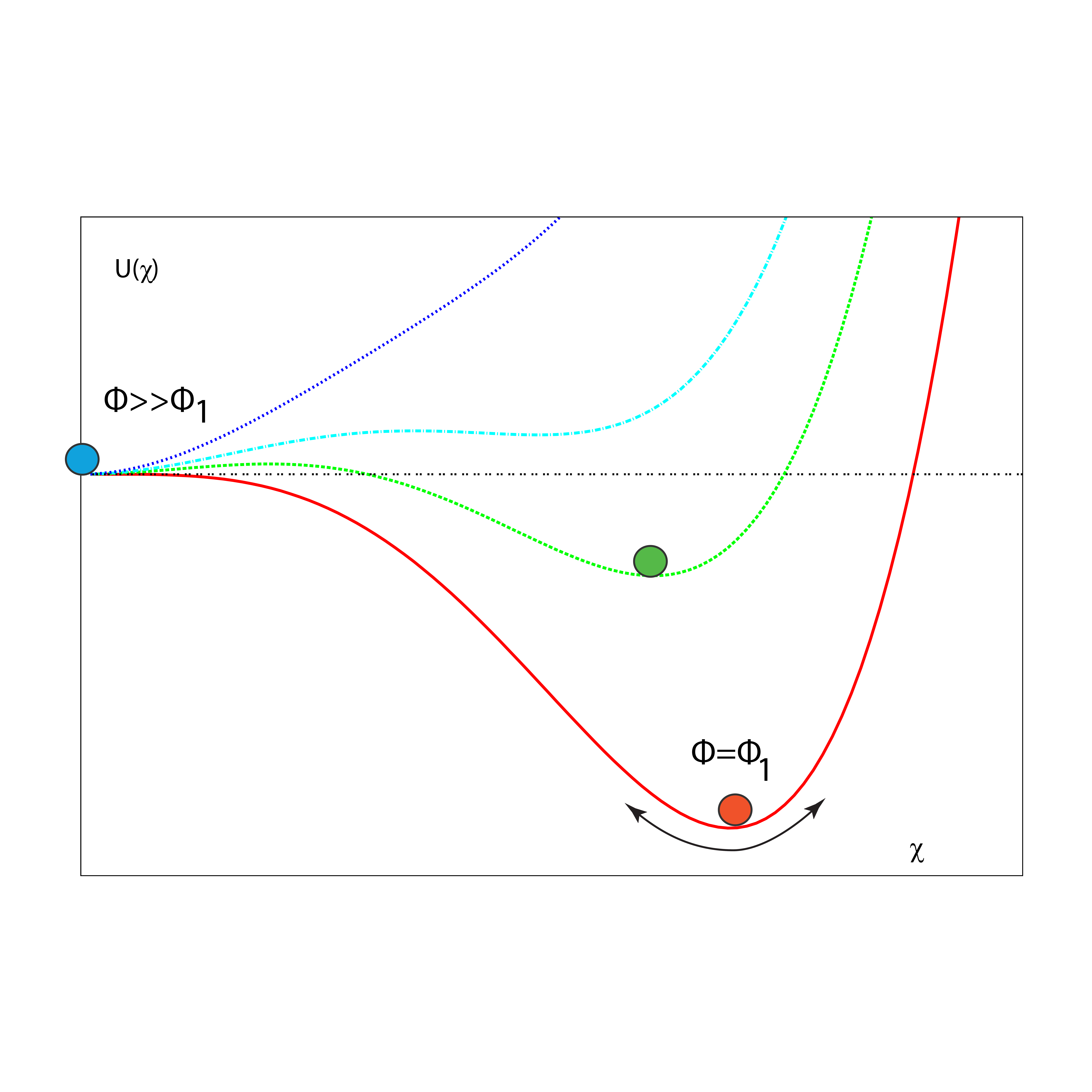}
\hfill
\includegraphics[width=0.48\textwidth]{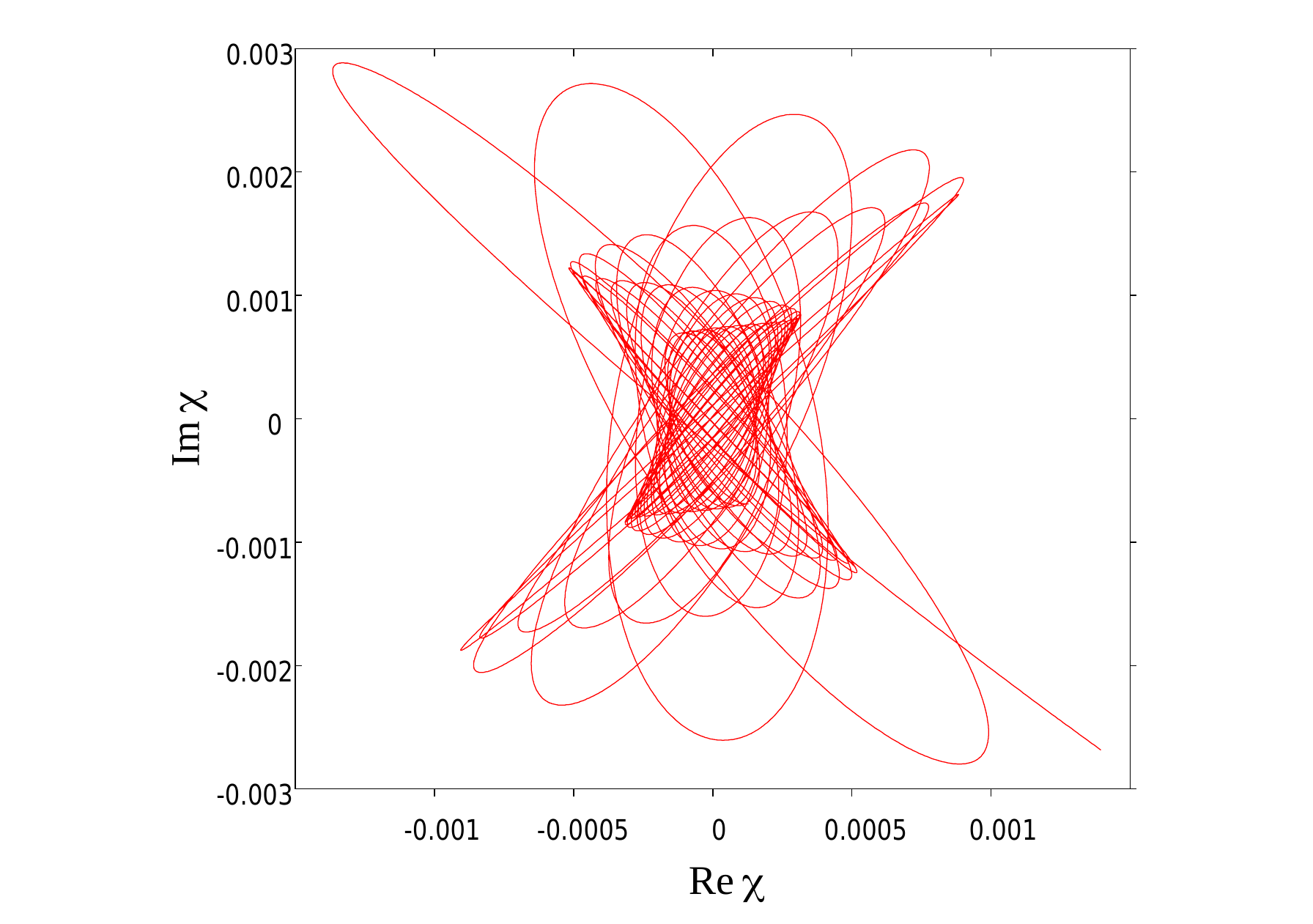}
\caption{Left: behavior of the effective potential of $\chi$ for different values of the inflaton field $\Phi$.
The upper blue curve corresponds to $\Phi \gg \Phi_1$ which gradually decreases to $\Phi = \Phi_1$, the red curve.
Then the potential returns back to an almost  initial shape, as $\Phi\to 0$. The evolution of $\chi$ in such
a potential is similar to the motion of a point-like particle (shown as the colored ball) in Newtonian mechanics.
First, due to quantum initial fluctuations, $\chi$ left the unstable extremum of the potential at $\chi = 0$ and
``tried'' to keep pace with the moving
potential minimum and later starts oscillating around it with decreasing  amplitude.
The decrease of the oscillation amplitude is due to the cosmological expansion.
In mechanical analogy, the effect of the expansion is equivalent to the liquid friction term,
$ 3H \dot \chi$. When $\Phi<\Phi_1$, the potential recovers
its original form with the minimum at $\chi = 0$, and $\chi$ ultimately
returns to zero, but before that it could give rise to a large baryon asymmetry \cite{ad-mk-nk}.
Right: The evolution of $\chi$ in the complex $[ {\rm Re} \chi, {\rm Im} \chi ]$-plane \cite{ad-mk-nk}.
One can see that  $\chi$ ``rotates'' in this plane with a large
 angular momentum, which  exactly corresponds to the baryonic number density of $\chi$.
Later $\chi$ decays  into quarks and other
particles creating a large cosmological baryon asymmetry.}
\label{f:ADBD}
% \end{center}
\end{figure}

Our model of early formation of heavy primordial black holes is based on the supersymmetric scenario of baryogenesis
(Affleck and Dine mechanism)~\cite{ad-bg}
slightly modified by an addition of general renormalizable coupling of the scalar
baryon field $\chi$ to the inflaton field $\Phi$, see figure~\ref{f:ADBD}.
The potential of the interacting fields $\Phi$ and $\chi$ are taken in the form:
\begin{equation}
U  (\chi, \Phi) = U_\Phi (\Phi) + U_\chi(\chi)
+ \lambda_1(\Phi-\Phi_1)^2|\chi|^2 ,
%+\lambda_2|\chi|^4\ln{\frac{|\chi|^2}{\sigma^2}}+ m_0^2 |\chi|^2
%+ m_1^2\chi^2+m_1^{*2}\chi^{*2}.
\label{U-of-chi-Phi}
\end{equation}
where the first term $U_\Phi (\Phi)$ is the potential of the inflaton field which ensures sufficient inflation,
$U_\chi(\chi)$ is the potential of the scalar field with non-zero baryonic number. According to the suggestion of
ref.~\cite{ad-bg}, this potential possesses flat directions along which the $\chi$ could travel away from the origin
accumulating a large baryonic number density. The last term in eq.~(\ref{U-of-chi-Phi}), introduced in
ref.~\cite{ad-js}, closes the flat directions when the inflaton field $\Phi$ is away from some constant value $\Phi_1$.
The latter is chosen
such that  $\Phi$ passes through $\Phi_1$ during inflation, close to its end when the remaining number of e-
foldings was about 30-40. This number depends upon the model can can be adjusted if necessary to comply with
specific observational requirements.

The potential (\ref{U-of-chi-Phi}) allows for an open window to the flat directions only during a relatively short period of time.
Correspondingly, the probability of the scalar baryon, $\chi$, to reach a large value is small because the initial evolution of $\chi$ from the
unstable minimum at $\chi = 0$ is a slow stochastic one, similar to the Brownian motion.
Most of  $\chi$ would remain small, and in these
regions of the universe the baryon asymmetry would have its normal value, $\beta = 6\times 10^{-10} $. However, there would be some
small bubbles where the asymmetry might be huge, even of order unity.
With properly chosen parameters of the model these bubbles
would have astrophysically interesting sizes  but
occupy a relatively small fraction of the whole volume of the universe. Despite that, due to large baryonic number,
they could make a noticeable
contribution to the total cosmological energy density.
The shape of the mass distribution of these high-B bubbles is determined by inflation and
thus it is virtually model independent
(except for the unknown values of the parameters) and has the log-normal form:
\begin{equation}
\frac{dn}{dM} = \mu^2 \exp\left[-\gamma\,\ln^2(M/M_{0}) \right],
\label{dn-dM}
\end{equation}
where $\mu$, $\gamma$, and $M_{0}$ are some unknown, model dependent constant parameters.
An interesting feature of this distribution is that a power-of-mass, $M^q$, factor instead of $\mu^2$
does not change the log-normal shape after the redefinition of parameters.

Originally, the induced small scale fluctuations of the baryonic number lead predominantly to isocurvature fluctuations, but
after the QCD phase transition at the temperature $T = 100 - 200 $ MeV, when the massless quarks turn into heavy
nucleons, the isocurvature fluctuations transformed into large density perturbations at
astrophysically large but cosmologically small scales.

Depending upon the history of their formation, the high-B bubbles could be either primordial black holes, compact stellar-like
objects, or even rather dense primordial gas clouds.
In particular at the high mass tail of the distribution superheavy PBHs might be created.
Their emergence at redshifts of order 10 is highly mysterious and even their existence in each
large galaxy is in serious tension with the conventional mechanisms of their formation.  An early formation of superheavy
black holes which later serve as seeds for galaxy formation looks more probable. For a review of the problems with rich
evolved population in the early universe see sec. 5 of ref.~\cite{ad-hi-z} or more recent review in the lecture~\cite{ad-itep-16}.
The accumulated data of the last few years suggest an early SMBH formation.

We can fit the parameters of the distribution (\ref{dn-dM})
%using the data on MACHOs~\cite{macho-all}
using the following three sets of data and/or assumptions:
\begin{itemize} %[A.]
\item[A] The fraction of dark matter in MACHOs is $f \approx 0.1$ for the mass range $(0.1 - 1) M_\odot$ (see, e.g. \cite{Bennett2005}).
\item[B] All primordial black holes make the whole cosmological dark matter
\item[C] The number density of the primordial black holes  with masses above $10^3 M_\odot$ is equal to the number
density of the observed large galaxies.
\end{itemize}

\begin{figure}[tbp]
\begin{center}
\includegraphics[scale=0.5]{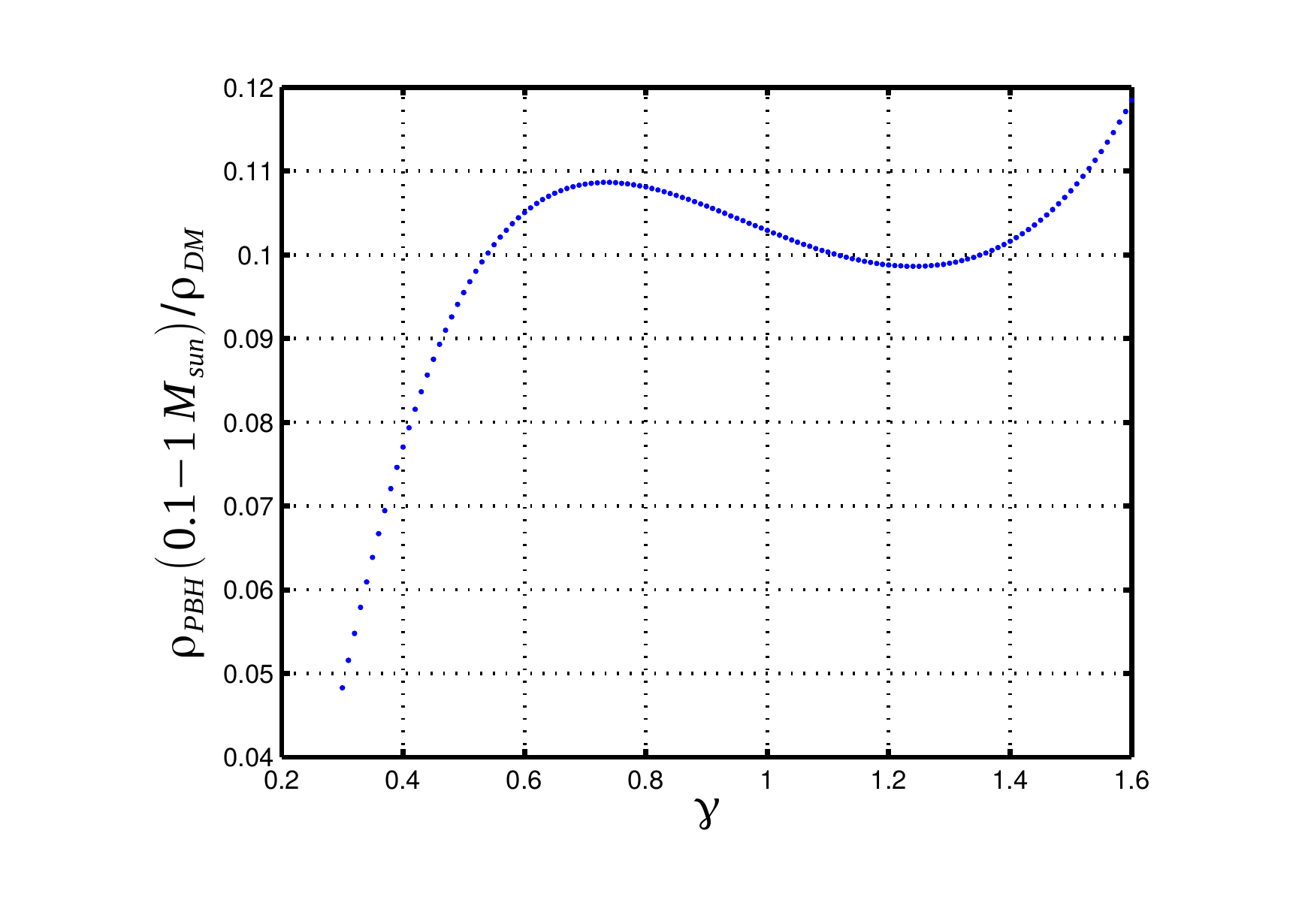}
\end{center}
\caption{Fraction of PBH density  ${\rho(0.1-1)/\rho_{DM}}$ in the mass range  $0.1 - 1 M_\odot$ as a function of parameter ${\gamma}$. See Eq. (\ref{dn-dM}).}
\label{S2}
\end{figure}

%Condition A is evident but the fraction $f$ can be somewhat smaller or larger by a factor of a few.
(A) The MACHO group has concluded that compact objects with masses
\newline
$ { 0.15M_\odot < M < 0.9M_\odot } $ have a fraction $f$ in the Galactic  halo
in the range ${ 0.08<f<0.50}$ (95\% CL).
Therefore, based on the results of MACHO group, it is recognized that  a population of compact lensing
objects really exists \cite{Bennett2005}.
There were some contradicting results based on microlensing observations of Andromeda galaxy M31,
see reviews in \cite{2014PhyU...57..183B,sb-ad-kp}.
A recent study of M31 \cite{2015ApJ...806..161L}  has discovered 10 new microlensing events
and the authors conclude: ``statistical studies and individual microlensing events
point to a non-negligible MACHO population, though the fraction in the halo mass remains
uncertain''. To fit this constraint, the parameter $M_0$ and $\gamma$ of the distribution \ref{dn-dM}
should be related as $M_0 = M_\odot (\gamma + 0.1 \gamma^2 - 0.2 \gamma^3)$
(see figure~\ref{S2}).

(B) Condition B is not obligatory
and can (must) be diminished in accordance with the bound on the BH dark matter.
Condition C should be fulfilled if all large
galaxies were seeded by a superheavy PBH. The initially formed superheavy PBH might have much smaller masses
(around $10^3-10^5 M_\odot$) which
subsequently grow to $10^{9} M_\odot$ because of an efficient accretion of matter and
mergings. This mass enhancement factor is
much stronger for heavier BH and
thus their mass distribution may be different from (\ref{dn-dM}). However, we assume that the original BHs were created with
the distribution (\ref{dn-dM}). Note that the accepted lower mass limit $M>10^3 M_\odot$ is rather arbitrary and is used here only
for the sake of simple parameter estimates.

Conditions A,B, and C can be crudely satisfied if the parameters in the initial PBH mass distribution
(\ref{dn-dM}) are
$\gamma = 0.5$, $M_{\max} =
M_\odot$, and $\mu = 10^{-43}$ Mpc$^{-1}$ in the natural system of units $\hbar=c=1$.
Here the value of $\mu$ is fixed by the condition that all Dark Matter is described by objects
with the distribution (\ref{dn-dM}).
The energy density of black holes with masses above $10 M_\odot$ in this case would be close to the energy density of the cosmological dark matter.

Let us note that the B-bubbles are formed predominantly spherically symmetric because such configurations minimizes
the bubble energy. Their angular momentum should be zero because they are formed as a result of phase transition in
cosmological matter with vanishing angular momentum
due to absence of the cosmological vorticity perturbations.
This is in striking contrast to BHs formed through the matter accretion,
when the angular momentum must be substantial.

\section{An upper limit on ADBD PBH.}
Let us discuss what could be the maximum mass of PBH in the AD scenario.
The mass of the bubbles with high baryonic number density, or High B Bubbles (HBB),
after their formation at the inflationary stage was equal to:
\begin{equation}
M_{\rm infl} = \frac{4\pi}{3} \rho l^3,
\label{M-infl-1}
\end{equation}
where $ l \sim (1/H) \exp ( H (t - t_{in}) )$, $H$ is the Hubble parameter during inflation, $t_{in} $ is the moment when the window
to the flat direction became open, i.e. the inflaton filed $\Phi$ started to approach  $\Phi_1$ from above.

The cosmological energy density at this stage was
\begin{equation}
\rho = \frac{3 H^2 m_{Pl}^2}{ 8\pi} = \mbox{const}\,.
\label{rho}
\end{equation}
Thus we obtain for the maximum value of the HBB mass at the end of inflation at $t=t_e$:
\begin{equation}
M_{\rm infl}^{(\max)} = \frac{m_{Pl}^2}{2 H} e^{3H (t_e -t_1)}\,.
\label{M-max-1}
\end{equation}

\noindent In the instant heating approximation, when the universe was heated up to temperature $T_h$, one finds
\begin{equation}
\rho = \frac{\pi^2}{30} g_* T_h^4,
\label{temp}
\end{equation}
where $g_*\sim 100$ is the number of particle species in thermal equilibrium plasma after the inflaton decay.

Taking all factors together, we find for the maximum value of the HBB mass at the end of inflation:
\begin{equation}
M_{\rm infl}^{(\max)} = \left(\frac{90}{32\pi^3 g_*}\right)^{1/2} \,\frac{ m_{Pl}^3}{T_h^2}   \, e^{3H (t_e -t_1)} \,.
\label{M-max-2}
\end{equation}
After the end of inflation the HBBs were filled by relativistic matter and thus their masses dropped down as
$T/T_h$ because the density of relativistic matter decreased as $1/a^4$, while the HBB size rose as $a$, i.e. its
volume grew up as $a^3$, where $a$ is the cosmological scale factor.
After the QCD phase transition at $T_{MD} \sim 100$ MeV the matter inside HBBs turns into nonrelativistic one,
and the total mass of HBB becomes constant.
A reasonable value of $T_h$ is about $10^{14}$~GeV, but the exact
magnitude is  of minor importance because the  necessary duration of inflation to create  HBB with the present day
value of the  mass  $M_{\rm today}$ depends only logarithmically on the temperature.
So we find for the maximum value of the HBB mass today:
\begin{equation}
M_{\rm today } =  \left(\frac{90}{32\pi^3 g_*}\right)^{1/2} \,\frac{ m_{Pl}^3}{T_h^2}\,\frac{T_{\rm MD}}{T_h}  \, e^{3H (t_e -t_1)}\,.
\label{M-today}
\end{equation}
Therefore, for the duration of inflation after the HBB formation which
is necessary to create PBH with the mass $M_{BH}^{ (\max) }$  we find:
\begin{equation}
e^{3 H (t_e - t_1)} = 10^{37 +5-10+15} \left( \frac{T_h}{10^{14} {\rm GeV}}\right)^3 \, \frac{M_{\rm BH}^{ (\max) }}{10^4 M_\odot}\,
\frac{100 \; {\rm MeV}}{T_{\rm MD} }\,.
\label{exp-time}
\end{equation}
Thus, finally  we need $H (t_e -t_1) = 36 $ to create a PBH distribution
with the  mass cut-off of $10^4 M_\odot$, if all other factors are taken to be unity.
Such a duration of inflation after  $\Phi = \Phi_1$ looks very reasonable.
In these estimates the Planck mass was taken to be $m_{Pl} = 10^{-5 } \mbox{g} = 10^{19} $~GeV.

In principle, we can turn equation (\ref{M-today}) another way around to find the high mass cut-off for fixed
duration of inflation after the HBB formation.

\section{Black hole growth.}
\label{BHgrowth}

The very first estimations of the accretion growth of
PBHs in radiation-dominated epoch were done by Zel'dovich and Novikov \cite{1967SvA....10..602Z}.
Using non-relativistic Bondi-Hoyle accretion analysis the authors have shown
that the masses of the formed at $t=t_0$ BHs change as
\begin{equation}
M=\frac{M_0}{1-\frac{KM_0}{t_0}(1-\frac{t_0}{t})},
\end{equation}
where $K=9\sqrt{3}G/2c^3$.
According to this formula, a PBH grows at the same rate as the Universe if its initial size is comparable
to the cosmological horizon, and at the end of radiation-dominated epoch the predicted masses of
the resultant BHs would become unrealistically high.
Zel'dovich \& Novikov method was generalized for the relativistic case in
\cite{2005PhRvD..71j4010H}, \cite{2004PhRvL..93b1102B}.
However, while accounting for cosmological decrease
of the density of the accreted substance, the previous simplified analyses ignored the influence of cosmological
expansion on the dynamics of accretion itself. This effect can be significant in the case of massive PBHs in
the early Universe when the BHs' radius can approach the Universe horizon. In our case,
the mass of PBH is
\begin{equation}
  M_0 \approx 10^5 M_\odot t
\end{equation}
for $t$ in seconds,
that is the Schwarzschild radius of PBH
\begin{equation}
r_{g} \simeq c t = 3\times 10^{10}  \frac{t}{\mbox{s}}\, \mbox{cm}
\end{equation}
is comparable to the size of horizon $r_{\rm H} \simeq 2ct = 6\times 10^{10} (t/\mbox{s})\, \mbox{cm} $.
Carr and Hawking \cite{1974MNRAS.168..399C} were the first who had drawn attention to downsides of
estimations based on the Bondi-Hoyle formula. Their results were analytically confirmed in
\cite{1978ApJ...219.1043B}, \cite{1978SvA....22..129N}. According to Carr-Hawking mechanism of
accretion, the growth of very massive PBHs in Zel'dovich \& Novikov analysis is significantly
overestimated \cite{2010CQGra..27r3101C}. However, Lora-Clavijo et al \cite{2013JCAP...12..015L}
numerically showed substantial growth of relatively smal$M_0 = M_\odot (\gamma + 0.1 \gamma^2 - 0.2 \gamma^3)$l PBHs ($r_g \ll r_{\rm H}$) during the
leptonic era (up to 100 sec).

\section{Massive PBH as galaxy seeds.}

The state-of-the-art SMBH growth paradigm, see e.g. resent EAGLE results
\cite{2016arXiv160400020R}, assumes a delta-like initial BH seed
masses of about $10^{5} M_\odot$, which formed in DM halos
at $z\sim 15$ and rapidly grow through gas accretion and
host galaxy mergings by $z\sim 5$. In the DM halos less massive than $\sim 10^{12} M_\odot$,
virtually no growth of the seed BH mass is observed, while at larger halo masses
the BH mass increases rapidly until a self-regulating BH mass -- DM halo mass
relation ($M_{\rm BH}-M_{200}$) is established.
The seed SMBH masses increase up to $10^8 M_\odot$, after which
their growth is linearly proportional to the surrounding DM halo mass.
 Importantly, the
shape of the SMBH mass function almost does not  change after $z\sim 5$, exhibiting
only the normalization increase by  an order of magnitude by $z=0$.
These finding are independent on the assumed seed BH mass for $M_{\rm seed}  \gtrsim 10^4 M_\odot$.

However, if there is a population of primordial SMBHs,
%with a given initial mass function,
they can serve as seeds for galaxy formation. Guided by
analogy with numerical simulations discussed above, we can assume that

1) seed BH masses $\lesssim 10^5 M_\odot$ do not substantially  evolve,
i.e. in the low mass interval the observed SMBH mass function has the primordial shape;

2) seed BH masses in the range $10^5-10^8 M_\odot$ rapidly grow up
to $\sim 10^8 M_\odot$;

3) seed BH masses exceeding $10^8 M_\odot$ at $z=0$ scale linearly with DM halo mass;

4) the rapid SMBH growth is completed by $z=5$ after which the SMBH mass function
(for $M>10^6$) linearly increases up to the present-day values.

Points (2) and (3) imply that the actual shape of the primordial SMBH mass function
is not very important for seed BH masses $\gtrsim 10^5 M_\odot$, since these BHs
rapidly grow to reach self-regulation in the surrounding DM halos.

\begin{figure}[tbp]
\begin{center}
\includegraphics[scale=0.5]{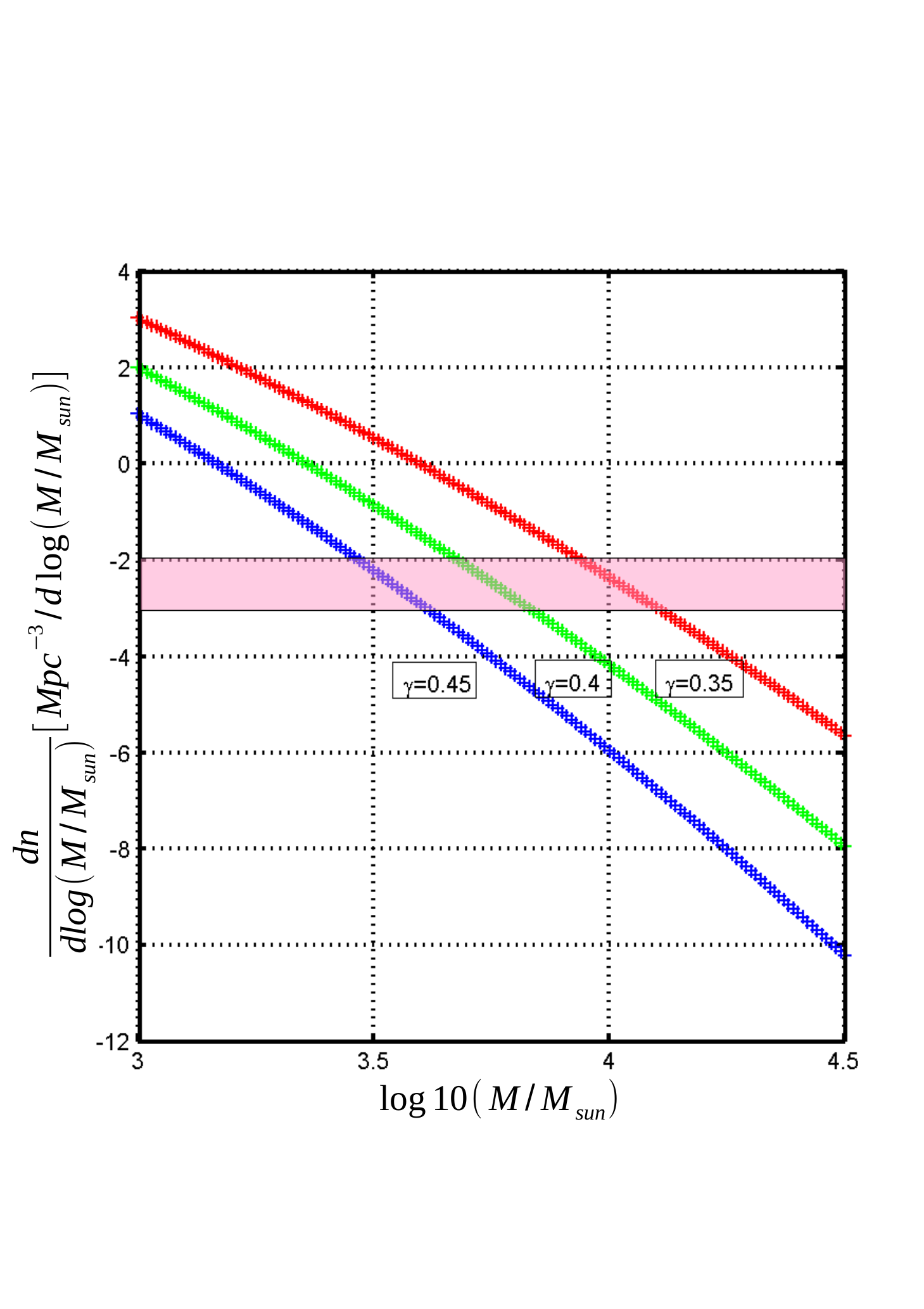}
\end{center}
\caption{Space density of PBH (per Mpc${^{-3}}$ per ${d\log M)}$, to be compared to those derived from observations  of SMBH in large galaxies (purple rectangle).}
\label{S3}
\end{figure}

Therefore, fitting the PBH mass function normalization
in the $10-100$~M$_\odot$ range to the BH+BH merging rate derived from the LIGO BH+BH detections  ($ 9-240$
events a year per cubic Gpc), we should only take care that the mass density of primordial SMBHs
does not contradict the existing SMBH mass function as inferred from observations of galaxies,
$dN/(d\log MdV)\simeq 10^{-2}-10^{-3}$~Mpc$^{-3}$ (see figure~\ref{S3}).

% \newpage

%From the review of BH growth at the radiation-dominated stage (see Supplementary material) we can conclude that large-mass part of the initial
%mass distribution of PBHs, created after the first second, will undergo minor changes, while PBHs
%with small masses can considerably grow in size.
%\textbf{PK: I don’t understand this – does it mean that the initial spectrum should be somehow transformed by z<3000 to be compared with GW Observations and SMBH density? If yes, how?}

%Various scenarios of BH formation from PopIII stars and subsequent formation of BH
%binaries are considered
%in \cite{2016arXiv160404288D}.
%They find the scenario \cite{2012ApJ...749...91F} as the most appropriate.
%This means that the contribution of PopIII stars in formation of BH binaries with masses
%$\sim 30+30 M_\odot $ is negligible, contradicting the population synthesis
%results \cite{Kinugawa2014,Kinugawa2016}.

%Therefore, by assuming the BH mass function normalization at the
%lower end to the BH+BH merging rate derived from the LIGO GW detection  ($\sim 9-240$
%events a year per cubic Gpc), we should only care that the higher SMBH mass density
%does not contradict the existing SMBH mass function as inferred from observations of galaxies,
%$dN/(d\log MdV)\simeq 10^{-2}-10^{-3}$~Mpc$^{-3}$.

\section{LIGO GW150914 as ADBD PBH event.}

\begin{figure}[tbp]
\begin{center}
\includegraphics[scale=0.65]{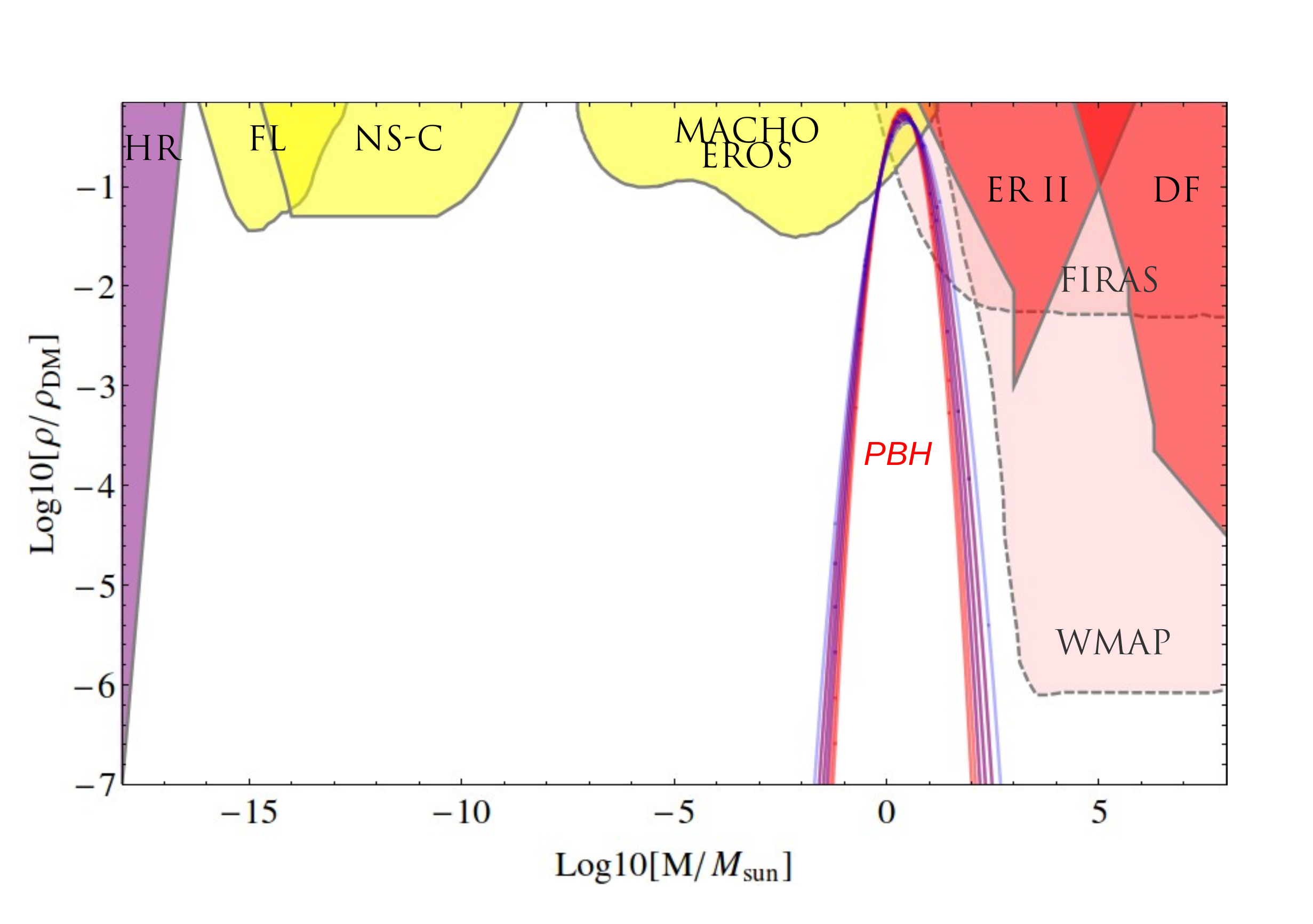}
\caption{Constraints on PBH fraction in DM, $f=\rho_{\rm PBH}/\rho_{\rm DM}$,
where the PBH mass distribution is taken as
$\rho_{\rm PBH}(M)=M^2dN/dM$ (see \cite{2016arXiv160706077C}).
The existing constraints (extragalactic $\gamma$-rays from evaporation (HR),
femtolensing of $\gamma$-ray bursts (F),
neutron-star capture constraints (NS-C), MACHO, EROS, OGLE microlensing (MACHO, EROS)
survival of star cluster in Eridanus II (E),
dynamical friction on halo objects (DF),
and accretion effects (WMAP, FIRAS))
are taken from \cite{2016arXiv160706077C}
and reference therein. The PBH distribution is shown for ADBD parameters $\mu=10^{-43}$~Mpc$^{-1}$,
$M_0=\gamma+0.1\times\gamma^2-0.2\times\gamma^3$ with $\gamma=0.75-1.1$ (red solid lines),
and $\gamma=0.6-0.9$ (blue solid lines).}
\label{f:PBH}
\end{center}
\end{figure}

Fixing the parameters ${\mu}$, ${\gamma}$, and ${M_0}$  in (\ref{mudn-dM}) we can
satisfy all existing constraints, understand the origin and properties of the merging BBHs, and explain other exciting puzzles.
We have found that the model parameters
${\mu = (2-3)\cdot 10^{-43} }$ Mpc$^{-1}$,
${\gamma = 0.4-1.0 }$,
and $M_0 = M_\odot (\gamma + 0.1 \gamma^2 - 0.2 \gamma^3)$ are suitable
to explain most of DM as ADBD BHs, do not contradict all existing constraints on PBHs \cite{2016arXiv160706077C}
and give sufficient amount of massive PBHs as seeds for early galaxy formation  (see figure~\ref{f:PBH}).
Note that CMB constraints from early accretion onto PBH
shown in the analogous plot \cite{2008ApJ...680..829R} are model-dependent and
should be treated with caution.
%

% \section{Creation of binary black hole systems}
% \label{BHcreaBin}

\begin{figure}[tbp]
\begin{center}
\includegraphics[scale=0.5]{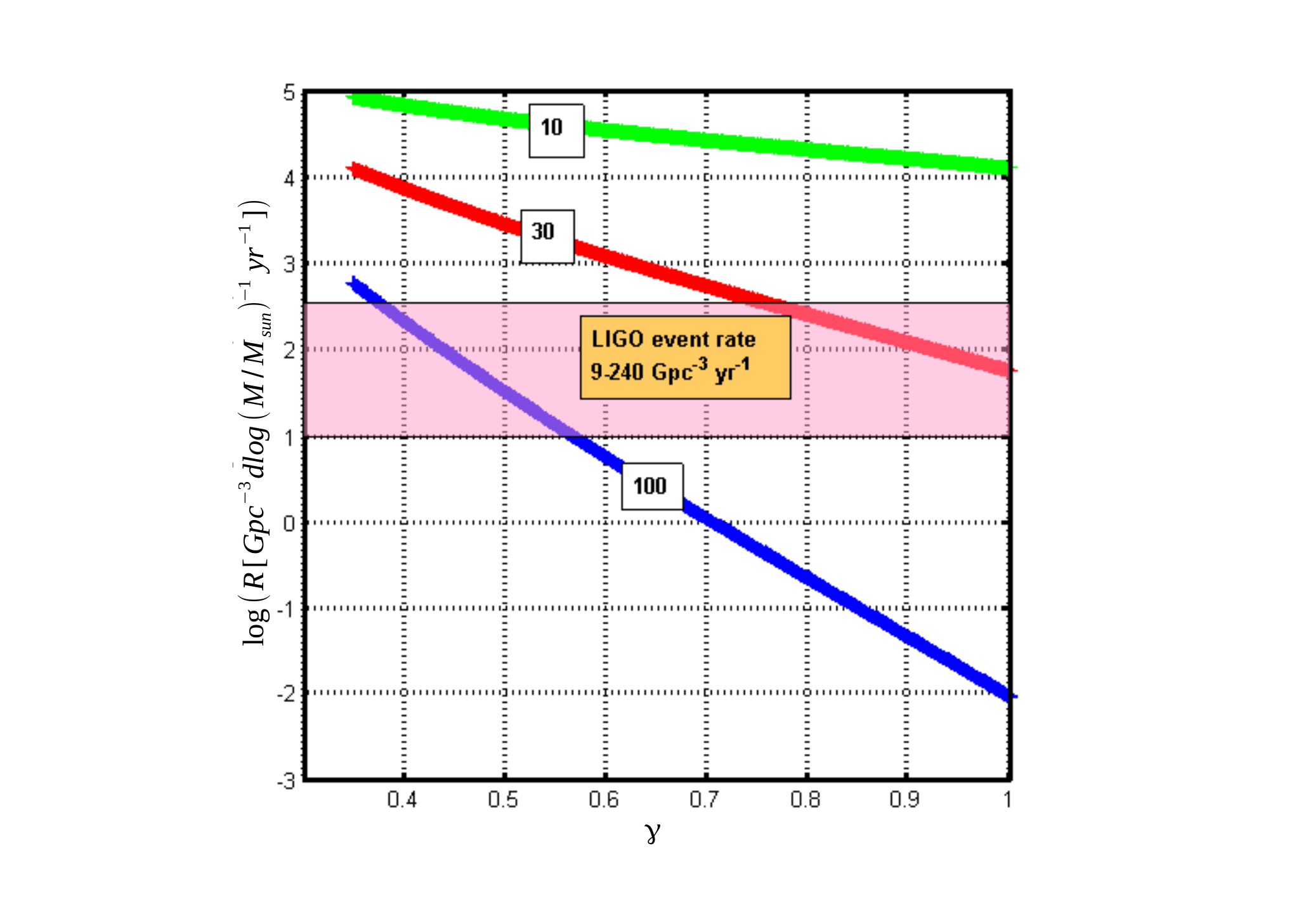}
\end{center}
\caption{Merging rate of binary BH $R$ (1/Gpc${^{3}}$/yr) per logarithmic mass interval over the universe age. The observed LIGO event rate is shown by the purple rectangle \cite{2016arXiv160203842T}.}
\label{S1}
\end{figure}

%Let us discuss now the question on formation of PBH binaries in AD-BD scenario.
Assuming the fraction of binaries among the PBH to be $10^{-3}$, the present-day merging rate
of binary BPH can be estimated as the space density divided by the Hubble time ($10^{10}$~years)
(see figure~\ref{S1}). Clearly, this can be easily made compatible with the binary BH
merging rate as inferred from LIGO observations (the purple rectangle in figure~\ref{S1}) by assuming
the parameters of the initial ADBD PBH distribution which are consistent with MACHO and SMBH
constraints discussed above.

Binary systems from PBHs started forming already at the radiation-dominated epoch \cite{1997ApJ...487L.139N}.
Assuming the population of PBHs with the same mass $M_{\rm BH}$, on the stage of matter-radiation equality with radiation density $\rho_{\rm eq}$ the average distance between BHs can be defined as:
\begin{equation}
\bar{x}=\left(\frac{M_{\rm BH}}{f\rho_{\rm eq}}\right)^{\frac{1}{3}},
\end{equation}
where $f$ is the fraction of DM in MACHOs. Two PBHs separated by the co-moving distance $x$
form a binary system and decouple from cosmological expansion
when the absolute value of mean gravitational energy of the PBH binary becomes higher than
the energy of the universe within horizon:
\begin{equation}
 a >  \frac{L_{\rm eq} }{f} \left(\frac{x}{\bar{x}}\right)^3,
\end{equation}
where $a$ is the scale factor,
and $L_{\rm eq} \sim  (3c^2/8\pi G \rho_{\rm eq})^{1/2} $
is the horizon scale at matter-radiation equality \cite{1997ApJ...487L.139N}.
Along with mean fluctuation field and angle
dependence of the tidal force, the authors of \cite{1998PhRvD..58f3003I},
have accounted for an important mechanism of 3-body interaction.
The tidal interactions
with neighboring PBHs provide the almost born binary system with sufficient angular
momentum to
prevent the two components of the binary from merging even if
PBHs are assumed to be born with negligibly small relative velocities.
All these effects should be accurately taken into consideration, since ignoring them leads to a $50\%$ uncertainty in the PBH binary merging rate estimates
\cite{1998PhRvD..58f3003I}.
%Moreover, the authors of \cite{1998PhRvD..58f3003I}
%have shown that

These estimates can be easily generalized
for the non-monochromatic mass spectrum  of PBHs presented in (\ref{dn-dM}).
%for the case of
%non-monochromatic mass spectrum  of PBHs like in (\ref{dn-dM}).
For PBHs with the mass $30 M_{\odot}$ the predicted event rate of GW bursts
\cite{2016arXiv160308338S} matches
the one observed by LIGO  only  for $f\simeq 7\times10^{-4}-7\times10^{-3}$.
Taking into account the tidal forces from adiabatic density perturbation of dark matter \cite{2016arXiv160404932E}, the valid range of
$f$ shifts to $f\simeq 2\times10^{-3}-2\times10^{-2}$. These limits on $f$ can be satisfied assuming different ranges of parameter $\gamma$
of log-normal distribution: $\gamma\simeq 0.75-1.1$ and $\gamma\simeq 0.6-0.9$ in the first and second case, respectively (see figures~\ref{S3} and \ref{S1}).

Binary PBHs can also form later in the matter-dominated epoch.
Driven by dynamical friction with collisionless particles,
PBHs can form gravitationally bounded pair in the centers of DM halos. However, the
effectiveness of this mechanism is poorly known and considerably model dependent \cite{1987gady.book.....B}.

\section{Discussion and conclusions.}

We have summarized the basic features of the HBB formation mechanism in figure~\ref{f:ADBD}.
In this scenario \cite{ad-js,ad-mk-nk} cosmologically small,
but possibly astronomically large, bubbles with high baryon density ${\beta}$ could be created in the early
universe, occupying a small fraction of the universe volume, while the rest of the universe has the standard
$\beta \approx 6\cdot 10^{-10}$.

The mass distribution of the high baryon density bubbles has the log-normal form:
\begin{equation}
\frac{dN}{dM} = \mu^2 \exp{[-\gamma \ln^2 (M/M_0)]},
% \nonumber
\label{mudn-dM}
\end{equation}
where ${\mu}$, ${\gamma}$, and ${M_0}$ are constant parameters.{\footnote{Though the log-normal mass distribution
is quite common for the usual stars, the considered here
 mechanism is the only one known to us for creation of such mass distribution of primordial black holes.}}
The shape of the spectrum is practically model-independent, it is determined by inflation.
There is an intrinsic high-mass cut-off in the spectrum set by duration of inflation
that can be $\sim 10^4$~M$_\odot$ or even higher (see  Section~\ref{BHgrowth}).
The constants in Eq.~(\ref{mudn-dM}) are not predicted by the theory, they can be fixed
from comparison with observations.
For example, the parameter $\mu$ can be determined by the normalization to the total number of high-B
bubbles (i.e. BD-stars and black holes) per comoving volume.

The formation of PBHs from the high-B bubbles proceeded in the following way.
After the QCD phase transition in the early universe at $ T = $ 100 - 200  MeV and
$ t \sim 10^{-5} - 10^{-4} $ seconds, quarks combined forming nonrelativistic baryons (protons
 and neutrons), thus  large density contrast arose between the bubble filled by heavy
baryons and the relativistic plasma in the bulk of the universe. When the bubble radius
reenters the cosmological horizon, the density contrast could easily be large,
$\delta \rho /\rho \gtrsim 1$, and thus the bubble
would turn into a BH. This mechanism is strongly different from those traditionally
considered in the literature (e.g. \cite{2014JCAP...01..037N} and references therein).

The promising constraints on PBHs density in several tens of solar masses range can be made with pulsar timing arrays through
investigation of third-order Shapiro delay \cite{2016arXiv161004234S}.
However, upper limit based on today's timing sensitivity will be within Eridanus and Firas estimates (figure~\ref{f:PBH}).

Thus, the PBHs formed in the AD-BD scenario possess the following properties:  \\
1. They have negligible  angular momenta (vorticity perturbations are practically absent). \\
 2. They have tiny peculiar velocities in the cosmological frame.\\
 3. They could be easily captured, forming binaries, due to dynamical friction in the dense early universe.\\
4. Their number density can be high enough to create an avalanche of
GW observations, especially with the higher sensitivity of the advanced LIGO and new GW detectors. \\
5. At the high-mass end of the universal log-normal PBH spectrum, they can seed an early SMBH formation in galaxies and even
seed the galaxy formation.

Measurement of BH mass and spin distribution from LIGO GW observations can be used to check the prediction of
the ADBD model presented here.

%Thus, the PBHs formed in the AD-BD scenario have the following properties:
%\begin{enumerate}
% \item
%{They have negligibly small angular momentum (vorticity perturbations are virtually absent).}
% \item
%{They have negligibly low peculiar velocity in the cosmological frame.}
% \item
%{They could be easily captured, forming binaries, due to dynamical friction in the dense early universe.}
% \item
%{Their number density can be high enough to create an avalanche of
%observed GWs, especially with the expected increase in sensitivity of the advanced LIGO and new GW detectors.}
%\item
%At high-mass end of the universal PBH spectrum, they can serve as seeds for early SMBH formation in galaxies.
%\end{enumerate}

%\textbf{Give citations to Ricotti \cite{2008ApJ...680..829R} on CMB limits from accretion
%onto PBH.}

%\textbf{Bird \cite{2016PhRvL.116t1301B} on GW from DM.}

%\textbf{Carr \cite{2016arXiv160706077C} PBH as DM with many results on accretion.}

% \subsection*{Additional figures}

\paragraph{Note added.}

S.B. is grateful to H.Niikura, M.Takada for valuable discussions on PBHs.
Additional historical and modern references are found in papers
\cite{1985qugr.conf..690K,2002astro.ph..2505K,2015PhRvD..92b3524C,2016Natur.534..512B,2016arXiv160802174C,Green2016,2016arXiv161008725W}.
% \cite{1985qugr.conf..690K} PBH as DM already discussed.
%
% \cite{2002astro.ph..2505K} Massive PBH as seeds of galaxies.
%
% \cite{2015PhRvD..92b3524C} The same, Bellido.
%
% \cite{2016Natur.534..512B} Massive low-metallicity rapidly rotating stars for BH-BH binary.
%
% \cite{2016arXiv160802174C} Low probability for massive PBH merger from CMB.
%
% \cite{Green2016} Microlensing and dynamical constraints on PBHs.
%
% \cite{2016arXiv161008725W} Constraint on the abundance of PBHs in dark matter.

\acknowledgments

A.D. acknowledges support of the Grant of President of Russian Federation
for the leading scientific Schools of Russian Federation,
NSh-9022-2016.2. S.B., N.P., K.P. acknowledge support from RSF grant 14-12-00203 (analysis of
PBH constraints from existing observations and LIGO GW150914 event).
The work of S.B. on manuscript was supported by World Premier International Research Center Initiative
(WPI Initiative), MEXT, Japan. N.P. acknowledges the support from IMPRS Bonn/Cologne and Bonn-Cologne Graduate
School(BCGS).

\bibliography{puzzlesBH}

\bibliographystyle{JHEP}

\end{document}